\documentclass[aps,prd,superscriptaddress,floatfix,preprint]{revtex4}

\usepackage[utf8]{inputenc}
\usepackage{graphicx}
\usepackage{tikz}
\usepackage{soul}
\usepackage{amssymb,amsmath,mathtools,comment,blkarray}
\usepackage{mathrsfs,array,tabularx}
\newcolumntype{C}[1]{>{\centering\let\newline\\\arraybackslash}p{#1}}

\usepackage{color} 

\newcommand{\ket}[1]{\left| #1 \right\rangle}
\newcommand{\bra}[1]{\left\langle #1 \right|}

\begin{document}

\title{Quantum computing with trapped ions: a beginner's guide}

\author{Francesco Bernardini}
\affiliation{Department of Electrical and Computer Engineering, University of Houston, Houston, Texas 77204-4005, USA}
\author{Abhijit Chakraborty}
\affiliation{Institute for Quantum Computing, University of Waterloo, Waterloo, ON, N2L 3G1, Canada}
\affiliation{Department of Physics, University of Houston, Houston, Texas 77024-5005, USA}
\author{Carlos R. Ord\'o\~nez}
\affiliation{Department of Physics, University of Houston, Houston, Texas 77024-5005, USA}

\date{\today}

\begin{abstract}
This pedagogical article elucidates the fundamentals of trapped-ion quantum computing, which is one of the potential platforms for constructing a scalable quantum computer. The evaluation of a trapped-ion system's viability for quantum computing is conducted in accordance with DiVincenzo's criteria. 
\end{abstract}

\maketitle

\setstcolor{red}
\section{Introduction} 
Two of the most interesting and promising active fields of research in the last 40 years, quantum computing (QC) and quantum information (QI), have taught us once more that even the hardest problems can be significantly simplified if we find a better way to reformulate them. Before the inception of QC and QI in the 1970s, the race to design the smallest integrated circuits and build the most powerful supercomputers was based on a \textit{classical} model of computing, where binary logic is the underlying language. After the seminal work of Feynman and others\cite{QChistory:1,QChistory:2,QChistory:3,QChistory:4,QChistory:5,QChistory:6,QChistory:7}, quantum computers have introduced a new language of computing based on quantum mechanics (QM), providing us with a totally novel way of simulating new problems and designing new algorithms with the ultimate goal of performing certain tasks faster than their classical counterparts. 

Despite being the hotbed of cutting-edge research, the basics of QC and QI are formulated using one of the simplest quantum models, the two-level system (for example, the familiar spin-1/2 system of the electron). This has a huge advantage from a didactic viewpoint as it allows interesting and useful activities to be developed even in an undergraduate environment \cite{edu:2,edu:3,edu:4,edu:5,edu:6}.

Nevertheless, QC and QI have potential for applications in many diverse areas of physical and social sciences like cryptography (more secure communications procedures using quantum principles) \cite{applications:1}, finance (quantum optimization algorithms to guide trading) \cite{applications:2}, quantum gravity  (simulations of black hole physics using quantum circuits) \cite{applications:gravity1,applications:gravity2}, and urban transportation (methods for intelligent traffic guidance using quantum computers)  \cite{applications:4}, to name a few.

Despite the significant advancements in quantum algorithms and communications theory, the biggest hurdle in the field remains the construction of a scalable quantum computer capable of performing these tasks. While there have been recent breakthroughs in building quantum machines \cite{QCmodern:1,QCmodern:2,QCmodern:3,qsupremacy}, we are still in the very early stages of QC, and there is no consensus within the scientific community regarding the best platform for realizing a quantum bit (qubit). However, over the years, trapped ion (TI) systems \cite{trappedions:1,trappedions:7,trappedions:8}, along with neutral atoms, \cite{neutralatoms-review} have emerged as one of the most viable QC platforms due to its scalability and ease of manipulation over other platforms like nuclei (using nuclear magnetic resonance) \cite{otherplatforms:1,otherplatforms:4,otherplatforms:5}, quantum optical devices  \cite{otherplatforms:2,otherplatforms:3}, superconductors \cite{qsupremacy,superQIrev} and quantum dots \cite{QIQdots}. Companies like Honeywell and IonQ have already made publicly usable quantum computers using trapped ions, and many research groups across the world also actively use trapped ion platforms for quantum simulations. While IBM has made significant advancements in using the superconducting platform for quantum computing, they are still limited by connectivity issues, making long-range connections between qubits error-prone. This is not a problem for trapped ions. However, trapped ion systems suffer from their own disadvantages, which we mention briefly at the end of this article. 

There already exists a vast amount of literature and books on trapped ion platforms   \cite{book:1,book:LesHouches,trappedionrev1,trappedionrev2}, but there is a lack of articles that explain the basics of quantum computing with emphasis on a physical realization of the system in an undergraduate environment. Scarani \cite{edu:1} in his paper explains nicely the basics of QC from a nuclear magnetic resonance (NMR) perspective. In this note, we use a similar pedagogical approach but emphasize the features of the trapped ions system that make them a good candidate for QC, instead of using a purely theoretical viewpoint. In this fashion, undergraduate students can also have an idea about how quantum computers can be practically realized and manipulated, much like semiconductor devices in classical computers. Connecting the theory with an experimental viewpoint is necessary to get a broader understanding of the physical intricacies and difficulties of building a quantum computer. Such aspects are often lacking in class-friendly literature, and our article aims to bridge this gap.

This article is organized in the following way: Secs. \ref{sec:qubits} and \ref{sec:universalgates} summarize the theoretical two-level system. In Sec.~\ref{sec:DiVincenzo}, we mention the qualities that a prospective QC platform must possess to build a successful quantum computer. In Secs. \ref{sec:ionprod}-\ref{sec:ion_readout} we concisely explain the mechanism of preparing a trapped ion system capable of performing quantum computations and how the laser-ion interaction can be used to change the ionic states. In Sec. \ref{sec:QIusingTI} we describe how TI systems can be manipulated to implement universal quantum gates, and finally provide some concluding remarks about the performance of TI systems in Sec. \ref{sec:Conclusion}.

\section{The qubit \label{sec:qubits}}
The building blocks of QC are called quantum bits (in analogy to the binary digit, or bit, of classical computing), or \textbf{qubits}. In quantum computing, a qubit is a two-state system where the states are typically eigenvectors of an observable. The physical meaning of this observable varies depending on the platform used to implement the qubit. In Dirac notation, the two states (orthonormal) are denoted by $\ket{0}$, and $\ket{1}$. A general state of the qubit can be in a superposition of the two states
\begin{equation}
\ket{q}=\cos\frac{\theta}{2}\ket{0}+e^{i\varphi}\sin\frac{\theta}{2}\ket{1}\;, \label{eq:qubitstate}
\end{equation} 
where $\theta\in[0,\pi]$ and $\phi\in[0,2\pi]$ are two parameters that define the state. 
The coefficients of $\ket{0},\ket{1}$ determine the probability to obtain the corresponding eigenvalues as a result of a measurement. Due to this probabilistic interpretation, the qubit state has unit norm (probabilities should add up to one), and the norm is usually preserved in any operation performed on the isolated system. These norm-preserving linear operations are usually called \textbf{unitaries}, or \textbf{gates} in QC. It can be shown \cite{book:1} that a generic unitary $U$ can be generated by suitable exponentials of the Pauli operators $\sigma_x$, $\sigma_y$, and $\sigma_z$, modulo a phase factor:
\begin{equation}
U=e^{i\alpha}e^{in_j\sigma_j\theta/2}\equiv e^{i\alpha}R_{\hat n}(\theta)\;,\nonumber
\end{equation}
\begin{equation}
\hat n=(n_x,n_y,n_z)\in \mathbb R^3,\quad\hat n\cdot\hat n=1\;,\label{eq:1qbgatedecomposition}
\end{equation}
where $R_{\hat n}(\theta)$ represents a rotation of angle $\theta$ around an axis pointed in the direction of $\hat{n}$. Pauli operators are Hermitian, traceless, but also unitary so they can be considered as gates themselves. They are defined by their commutation and anticommutation relations:
\begin{equation}
\left[ \sigma_i,\sigma_j \right]=2i\epsilon_{ijk}\sigma_k,\quad\left\{ \sigma_i,\sigma_j \right\} =2\delta_{ij}\,. \label{eq:paulicommutator}
\end{equation}
In the following, we will call $\left\{ \left| 0 \right\rangle,\left| 1 \right\rangle \right\}$ as the \textit{computational basis} and we will identify it with the set of eigenstates of $\sigma_z$:
\begin{equation}
\sigma_z \ket{0} = \ket{0} , \quad \sigma_z \ket{1} = -\ket{1}\;. \label{eq:sigmaz}
\end{equation}
The choice of this basis produces a matrix representation for the qubit state and operators:
\begin{gather}
 \left| 0 \right\rangle \coloneqq \begin{pmatrix}
1 \\0 
\end{pmatrix}
,\quad \left| 1 \right\rangle \coloneqq \begin{pmatrix}
0 \\1 
\end{pmatrix}
\; ,
\label{eq:vectors}\\
\sigma_x\coloneqq\begin{pmatrix}
0 & 1\\1 & 0 
\end{pmatrix}\,,\,\sigma_y\coloneqq\begin{pmatrix}
0 & -i\\i & 0 
\end{pmatrix}\,,\,\sigma_z\coloneqq\begin{pmatrix}
1 & 0\\0 & -1 
\end{pmatrix}\; . \label{eq:operators}
\end{gather}

\section{Multiple-qubit gates\label{sec:universalgates}}
For a $N$-qubit system, the state vector space (Hilbert space) is a tensor product of $N$ single-qubit spaces. For example, a general 2-qubit system can be in a superposition of $\ket{00},\,\ket{01},\ket{10},\,\ket{11}$ vectors. The notation $\ket{ab}$ here means the tensor product of the two one-qubit states $\ket{a}\otimes \ket{b}$, where $\ket{a}$ is the state of qubit 1 and $\ket{b}$ is the state of qubit 2. The gates acting on an N-qubit system are consequently $2^N\times 2^N$ unitary matrices generated from tensor products of 1-qubit gates. For $N=2$, an important family of gates is represented by the \textit{controlled gates}: according to the state of a \textit{control} qubit, a controlled gate performs a specific action on a \textit{target} qubit. One of the simplest examples is the $CNOT$, or controlled-$NOT$, acting on a pair $\ket{c}\ket{t}$ where the first qubit represents the control and the second the target:
\begin{itemize}
    \item If the control qubit $\ket{c}$ is in the $\ket{0}$ state, $CNOT$ does nothing on the target $\ket{t}$;
    \item If the control qubit $\ket{c}$ is in the $\ket{1}$ state, $CNOT$ applies $\sigma_x$ on the target $\ket{t}$.
\end{itemize}
The usefulness of this gate lies in the fact that the action of an arbitrary N-qubit gate can be mimicked with arbitrary accuracy using only 1-qubit rotations $R_{\hat n}(\theta)$ and $CNOT$ gates \cite{QChistory:7}. For this reason, if we can manipulate a physical qubit to perform these two operations, we can in principle perform any other complicated gates.

\section{Qubit entanglement}
One of the most distinctive features of QM, which is also crucial to implementing QC operations, was labeled by Einstein as a "spooky action at distance" after the formulation of the famous Einstein–Podolsky–Rosen (EPR) paradox \cite{Einstein1935}: such a feature is now commonly known with the name of \emph{entanglement}. Satisfactory treatment of entanglement goes beyond the scope of the present work, but the interested reader can consult one of the many reviews on the topic, such as \cite{horodecki2009}. In a nutshell, we can define entanglement as the  quantum correlation between multiple qubits, i.e., the effects of the correlation cannot be explained by the laws of classical physics. Due to this quantum correlation, it is not possible to write the entangled state of two qubits as the tensor product of two one-qubit states on any basis. To further illustrate the characteristic of an entangled state, we consider an example of a 2-qubit state of the form
\begin{equation}
    \ket{\Psi}=(\alpha_1\ket{0}_1+\beta_1\ket{1}_1)\otimes(\alpha_2\ket{0}_2+\beta_2\ket{1}_2)\;.
\end{equation}
where the explicit subscripts have been added (only in this section) to make a distinction between the two qubits. This state can be expanded in the sum of the various factors:
\begin{equation}     \ket{\Psi}=\alpha_1\alpha_2\ket{0}_1\ket{0}_2+\alpha_1\beta_2\ket{0}_1\ket{1}_2+\beta_1\alpha_2\ket{1}_1\ket{0}_2+\beta_1\beta_2\ket{1}_1\ket{1}_2\;.
\end{equation}
The statistics of the state are contained in the four parameters $\alpha_1,\beta_1,\alpha_2,\beta_2$, and if we measure the state of the first qubit, we collapse it in one of the two possible states $\ket{0}_1$ and $\ket{1}_1$ with probabilities $|\alpha_1|^2$ and $|\beta_1|^2$, respectively. However, the measurement operation on the first qubit has no effect on the second qubit. To see this, we recall the mathematical description of a projective quantum measurement, by means of the action of a (Hermitian) projection operator. In this case, if we assume that the outcome is the state $\ket{0}_1$, the projection operator will be $\mathcal P^{\ket{0}}_1=\ket{0}_1\bra{0}_1$, and we have:
\begin{equation}
\mathcal P^{\ket{0}}_1\ket{\Psi}=\ket{0}_1\bra{0}_1\left(\alpha_1\alpha_2\ket{0}_1\ket{0}_2+\alpha_1\beta_2\ket{0}_1\ket{1}_2+\beta_1\alpha_2\ket{1}_1\ket{0}_2+\beta_1\beta_2\ket{1}_1\ket{1}_2\right)\;,
\end{equation}The effect of this operator is to project the first qubit onto the subspace $\ket{0}_1$ ($\bra{0}_1\ket{0}_1=1$ and $\bra{0}_1\ket{1}_1=0$):
\begin{equation}
    \mathcal P^{\ket{0}}_1\ket{\Psi}=\alpha_1\alpha_2\ket{0}_1\ket{0}_2+\alpha_1\beta_2\ket{0}_1\ket{1}_2\equiv \alpha_1\ket{0}_1(\alpha_2\ket{0}_2+\beta_2\ket{1}_2)\;.
\end{equation}
Now, if we measure the second qubit on the state $\mathcal{P}_1\ket{\Psi}$, we can have as outcomes $\ket{0}_2$ and $\ket{1}_2$ with probabilities $|\alpha_2|^2$ and $|\beta_2|^2$, respectively, which is the same result we would have expected by conducting a computational basis measurement on the second qubit in state $\ket{\Psi}$, before performing the measurement on the first qubit. A state for which this happens is called \emph{separable} or \emph{factorable} and is by definition not entangled.
On the other hand, let us consider the state:
\begin{equation}
    \ket{\Phi} = \frac{1}{\sqrt2}\left(\ket{0}_1\ket{0}_2 + \ket{1}_1\ket{1}_2\right)\;,\label{eq:bellproto}
\end{equation}
Now, we perform a measurement on the first qubit and we assume again that the outcome is $\ket{0}_1$, so that the measurement operation is given by 
\begin{equation}
    \mathcal P^{\ket{0}}_1\ket{\Phi}=\ket{0}_1\bra{0}_1\frac{\ket{0}_1\ket{0}_2+\ket{1}_1\ket{1}_2}{\sqrt2}=\frac{\ket{0}_1\ket{0}_2}{\sqrt2}\;.
\end{equation}
A measurement on the second qubit, now, will yield the state $\ket{0}_2$ with probability 1. In other words, the first measurement not only collapsed the first qubit to $\ket{0}_1$, but also determined the outcome of any measurement done on the second qubit. A state $\ket{\Phi}$ for which this happens is called \emph{entangled}, and the property of two qubits being correlated via this purely quantum mechanism is called entanglement.

States of the form (\ref{eq:bellproto}) has a special characteristic. It turns out that the entanglement content of this state is the maximum (which can be quantified by a quantity called entanglement entropy). There are four states in total with this property in the Hilbert space of two qubits, which are called the Bell states (after the Irish physicist John S. Bell), also known as maximally entangled state:
\begin{equation}
    \ket{\Phi^+}=\frac{\ket{00}+\ket{11}}{\sqrt2},\,\ket{\Phi^-}=\frac{\ket{00}-\ket{11}}{\sqrt2},\,\ket{\Psi^+}=\frac{\ket{01}+\ket{10}}{\sqrt2},\,\ket{\Psi^-}=\frac{\ket{01}-\ket{10}}{\sqrt2},\;.
\end{equation}
Entangling qubits is a necessary feature to realize meaningful QC applications \cite{qc:entanglement}, including Shor's famous factoring algorithm and quantum teleportation to name a few. So, it is crucial to be able to create entangled states in a potential QC platform. We will describe how Bell states are created in a trapped ions system later in this article.
\section{Realization of qubits: DiVincenzo's criteria \label{sec:DiVincenzo}}
Any physical system whose mathematical description is satisfactorily formulated in terms of a two-level system is a good candidate to represent a qubit. As mentioned in the introduction, nuclei, optical systems, superconducting circuits, quantum dots, neutral atoms, and trapped ions are some notable examples. Different platforms have different advantages and disadvantages, and in order to benchmark their fitness for QC applications, DiVincenzo \cite{DiVincenzo} developed five criteria.
\begin{enumerate}
\item (a) Qubits must be well characterized and easy to produce. (b) the system must be scalable, i.e. must be able to manage an arbitrary number of qubits.
\item It must be possible to reliably initialize a qubit to a fiducial state, e.g. $\ket{0}$;
\item A qubit must be stable over timescales larger than the typical time needed to operate on it. The lifespan of a qubit over which it can maintain its quantum information content (before noise starts to destroy it) is known as coherence time. For a viable QC system, the coherence time of the qubit must be greater than the computation time. 
\item It must be possible to implement a universal set of quantum gates.
\item A reliable procedure must exist to read out the state of a qubit. 
\end{enumerate}
In this paper, we will focus on trapped ions, and describe their viability as a QC platform according to these criteria.

\section{Ion production \label{sec:ionprod}}
An ion is a nucleus or a molecule whose electron cloud has been deprived (cations) or augmented (anions) by one unit, so the total system is not electrically neutral. Among the several techniques available to (positively) ionize an atom \cite{ionization:4,ionization:5,ionization:6}, we will briefly mention photoionization \cite{ionization:1,ionization:7}. It consists of hitting the neutral atom with light tuned to a suitable wavelength, which is capable of exciting an outer electron and provide it with enough energy to leave the atomic orbital. Right after ionization, newly created ions are moved into a protected environment that delays electron reabsorption, for an amount of time sufficient to perform all the required experimental activities. However, the particular details of the procedure are not of primary interest to this review and we refer the interested readers to \cite{ionization:2,ionization:3} for further reading.

For QC applications, the choice of the element to ionize depends mainly on the atomic structure, and on the ionization energy. Elements belonging to the groups IIA ad IIB of the periodic table:
\begin{equation}
\text{Be, Mg, Ca, Sr, Ba, Zn, Cd, Hg, Yb}\nonumber
\end{equation}
are the most favorable in this sense and are therefore more commonly used.

After ionization, such elements will show a hydrogen-like spectrum of energy states $\ket{\mathfrak a_i}$, $i\in N$, characterized by energy levels $E_{a_i}$, eigenstates of an atomic Hamiltonian $H_{a}$. Two of these states, which we will call $\ket g$, \textit{ground}, and $\ket e$, \textit{excited}, will be singled out and identified with the abstract states $\{\ket0,\ket1\}$ of (\ref{eq:sigmaz}). 

The transition $\ket{g} \rightarrow \ket{e}$ can be chosen either in the optical or in the hyperfine range by selecting the ground and excited state properly from the ionic orbitals. The corresponding qubits are called \textit{optical} and \textit{hyperfine qubits}, respectively \cite{trappedions:7,trappedions:8}. See Table~\ref{tab:lifetimes} for a succinct comparison between the two. As we can see, the lifetimes (and hence, the coherence times) of the hyperfine qubits are much larger than the optical qubits, which usually makes it the preferred choice for QC, along with the fact that it is also easier to manipulate.

The current state of the art of ion manipulation techniques\cite{trappedions:7, iontechnique:new} allows rates of operation of hundreds of kHz, which means that basic operations over qubits can be performed in times much smaller than the typical lifetimes of optical or hyperfine qubits \cite{conclusions:3}. Therefore, optical and hyperfine qubits provide a stable, well-characterized, and rapidly manipulable platform, thus satisfying criteria 1a and 3 mentioned in Sec.~\ref{sec:DiVincenzo}. 

 \begin{table}
 \centering
    \caption{Typical parameters of the ionic optical and hyperfine transitions used in QC. In the examples $F, \,m_F$ represent total angular momentum including the isospin and its $z$ component.\\}
    \begin{tabularx}{\linewidth}{
    |>{\hsize=.6\hsize}X|
    >{\hsize=1.2\hsize}X|
    >{\hsize=1.2\hsize}X|
  }    
    \hline
     Properties & Optical qubits & Hyperfine qubits\\
     \hline \hline Range & visible, 380-740 nm, 405-790 THz & microwave, 3-300 mm, 1-100 GHz\\
     \hline Lifetime & $\sim$ 1 s & $\sim$ 10 min \\
     \hline States & $S$ level and a meta-stable state & Two hyperfine levels.\\
     \hline Example & $6S_{1/2} \equiv \ket{g}$ and $5D_{5/2}$ $\equiv \ket{e}$ levels of $\rm Ba^{+}$ & ${}^2S_{1/2}(F=1,m_F=0)$ $\equiv \ket{g}$ and ${}^2S_{1/2}(F=0,m_F=0) \equiv \ket{e}$ levels of $\rm Cd^{+}$\\
     \hline \hline 
    \end{tabularx}
\label{tab:lifetimes}
\end{table}

\section{Ion traps \label{sec:trapping}}
What we have described so far only concerns the ion's \textit{internal structure}. However, an ion also moves in space, therefore its total Hamiltonian $H$ will also feature a kinetic and a potential term:
\begin{equation}
H=H_{a}+\frac{\vec P^2}{2M}+V(\vec x)\equiv H_{a}+K+V\label{eq:totalenergy}\;,
\end{equation}where $\vec P$ is the momentum, $M$ is the mass of the ion and $V(\vec x)$ is the potential energy the ion is subjected to.

A newborn photoionized ion has typically an energy of $\sim$ 1 keV, which corresponds to a speed of $\sim$ 1 m/s for an ion such as $^{40}Ca^+$. At such energies, the ion can be considered a classical system from a motional perspective. So, in order to build a fully quantum system we need to constrain its motion both in range (\textbf{trapping}) and speed (\textbf{cooling}).

An \textbf{ion trap} is a device whose purpose is to keep ions confined within a narrow region of space: necessary condition for this to happen is that the ion feels a \textbf{minimum} of the potential energy $V(\vec x)$ (the \textbf{trap center}) at some point inside the trapping region. 

Taking advantage of the charged nature of the ions, a trapping potential can be created using a suitable combination of electric and/or magnetic potentials. Two paradigms exist for the design of such trapping potential:
\begin{itemize}
\item The Penning trap \cite{Penning:1,Penning:2}, where \textbf{static} electric \textbf{and} magnetic potentials are used;
\item The Paul trap \cite{Paul:1,Paul:2,Paul:3,Paul:4}, where \textbf{static and oscillating} electric fields are used. 
\end{itemize}
Although both are in principle suitable for QC applications, the Paul trap is currently the most widely used method in academic and industrial environments. 
\begin{figure}
    \centering
    \includegraphics[width=0.65\linewidth]{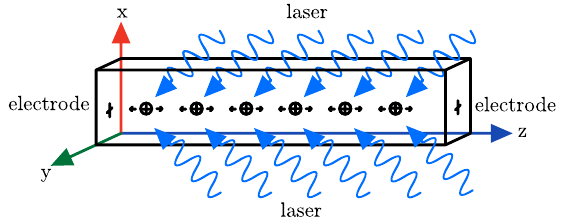}
    \caption{Ions strongly trapped by a Paul trap in the $xy$ plane, and allowed to move only in the $z$ direction. A pair of electrodes prevent the ions from repelling each other out of the trap.}
    \label{fig:trapped}
\end{figure}
A famous theorem due to Earnshaw \cite{earnshaw} states that a charged particle cannot find stable equilibrium in a static electric potential that satisfies the Laplace equation. However, it is possible to show that in the case of potentials rotating at particular frequencies, solutions of the equations of motion corresponding to a dynamical equilibrium point arise, allowing the charged particle to be confined in a desired volume. A gravitational analogue of a Paul trap can be seen in \cite{paultrap}.

Sufficiently close to the trap center, any potential can be well approximated by a quadratic function of the coordinates, i.e., a trapped ion will behave like a \textbf{simple harmonic oscillator} (SHO) in 3 dimensions. 
The most commonly trapping systems used in quantum computing applications are \textbf{linear traps}: the trapping force in two of the three dimensions is much stronger than the one in the third (for example, a very strong Paul trap in the $x$ and $y$ direction and a relatively weaker pair of electrodes in the $z$ direction, as in Fig.~\ref{fig:trapped}), therefore the only relevant movement will be a simple harmonic motion in (say) the $\hat z$ direction, characterized by a frequency $\omega_z$ and by the ion's mass $M$. Eq. (\ref{eq:totalenergy}) can be recast as 
\begin{equation}
H=H_{a}+\underbrace{\frac{P_z^2}{2M}+\frac12M\omega_z^2z^2}_{H_{SHO}}\equiv H_{a}+K_z+V_z\label{eq:SHOz}\;,
\end{equation}where $P_z$ is the momentum in the $\hat z$ direction. Typical experimental values for $\omega_z$ are in the order of $\sim$ 1 MHz.

In order for the harmonic approximation to hold, the ion must not move appreciably away from the trap center. If this happens, we cannot guarantee anymore that the forces acting on the ion are directed towards the center of the trap, and at some point it could even find itself outside of the trap.

For the above reason it is important that the ion moves slowly, so we want to lower its kinetic energy. With a slight abuse of language we call this stage ``cooling" by associating a temperature with the kinetic energy, even though the ion is not in thermal equilibrium. In Sec.~\ref{sec:laser_cooling}, we will describe two examples of \textbf{laser cooling}, valid in the classical and quantum regimes, respectively. Using suitably tuned lasers, it is possible to impart an effective damping force on the ions, which slows them down.

Finally, collisions with other particles are also detrimental to the trapping procedure as they can induce sudden changes in the kinetic energy, which may lead to the escape of ions from the trap. In order to minimize such interactions, an \textbf{ultra-high vacuum} \cite{vacuum:1,vacuum:2,vacuum:3} at pressures of about $10^{-11}$ Torr ($\sim1.3\times 10^{-9}$ Pa) is created in the region where the trap will be placed. 

\section{Laser cooling of ions \label{sec:laser_cooling}}
As mentioned in the previous section, we can identify two regimes in cooling depending on how the average oscillator energy $\langle H_{SHO}\rangle$ compares with the characteristic energy of the quantum SHO (QSHO), $\hbar\omega_z$:
\begin{itemize}
    \item A \textbf{classical} regime, where the ion can be considered a classical spring, and quantum phenomena can be neglected. This is a good approximation as long as $\langle H_{SHO}\rangle$ is much larger than $\hbar\omega_z$;
    \item A \textbf{quantum} regime, where conversely $\langle H_{SHO}\rangle$ is comparable to $\hbar\omega_z$, and quantum effects become important.
\end{itemize}

Our goal is to bring the ions down to the ground state of the QSHO, which can be used as the fiducial qubit or $\ket{0}$ on which computation can be performed (criterion 2 in Sec.~\ref{sec:DiVincenzo}). This is usually a two step process: (a) cooling the system to the point where the vibrational QSHO degrees of freedom (dof) become active, and (b) bringing the QSHO to its ground state. We describe the most commonly used methods to achieve this in the following two subsections.  

\subsection{Classical regime: Doppler cooling \label{sec:doppler_cooling}}
An ion is usually in the classical regime when produced. The primary method used to cool it down to the quantum regime is known as Doppler cooling \cite{dopplercooling:1}, which uses Doppler shift as the underlying cooling method. As a first step, a resonant transition between two states of the ion is chosen with frequency $\omega_0$ and linewidth $\Gamma$, which represents the width of the absorption maxima in the frequency space. $\Gamma$ is usually much larger than the QSHO frequency $\omega_z$, so that the QSHO levels remain unresolved. These ions are irradiated with a monochromatic laser with frequency tuned to a value slightly lower than the transition frequency: $\omega_{\rm abs} = \omega_0 - \delta\omega$ and wave vector $\vec{k}$ (momentum $\hbar \vec{k}$). Now consider the scattering of an ion moving with velocity $\vec{v}$ with the laser photon. Suppose that the ion absorbs the photon (the condition of which we will derive shortly), goes to an excited state and then spontaneously decays back by emitting a photon in a random direction with the same energy. For an absorption process, we can write the energy and momentum conservation equation as
\begin{gather}
    \hbar \omega_0 + \frac{1}{2} M v'^2 = \hbar\omega_{\rm abs} + \frac{1}{2} Mv^2\;, \label{eq:energy_conservation}\\
    M \vec{v'} = M\vec{v} + \hbar \vec{k}\;,\label{eq:momentum_conservation} 
\end{gather}
where $\vec{v'}$ is the velocity of the ion just after the absorption, $M$ is the mass of the ion, and the subscript `abs' represents the absorption process. Substituting $\vec{v'}$ from Eq.~(\ref{eq:momentum_conservation}) into Eq.~(\ref{eq:energy_conservation}) we get
\begin{equation}
    \omega_0 = \omega_{\rm abs} - \vec{v} \cdot \vec{k} - \frac{\hbar k^2}{2M} \;.\label{eq:doppler_absorption}
\end{equation}
On the RHS, the second term is the well-known Doppler shift, and the third term is known as the
recoil shift, which is usually small compared to the Doppler shift at high velocities. Neglecting the third term and inverting the expression, we get $\omega_{\rm abs} = \omega_0\, + \,\vec{v}\cdot\vec{k}$. Since the laser frequency was tuned to a value lower than the transition frequency $\omega_0$, absorption is only possible if $\vec{v}$ and $\vec{k}$ are in opposite directions (making $\vec{v} \cdot \vec{k}$ negative). Due to the Doppler shift, the ions moving towards the laser absorb the photon to go to an excited level.
\begin{figure}
    \centering
    \includegraphics[width=0.6\linewidth]{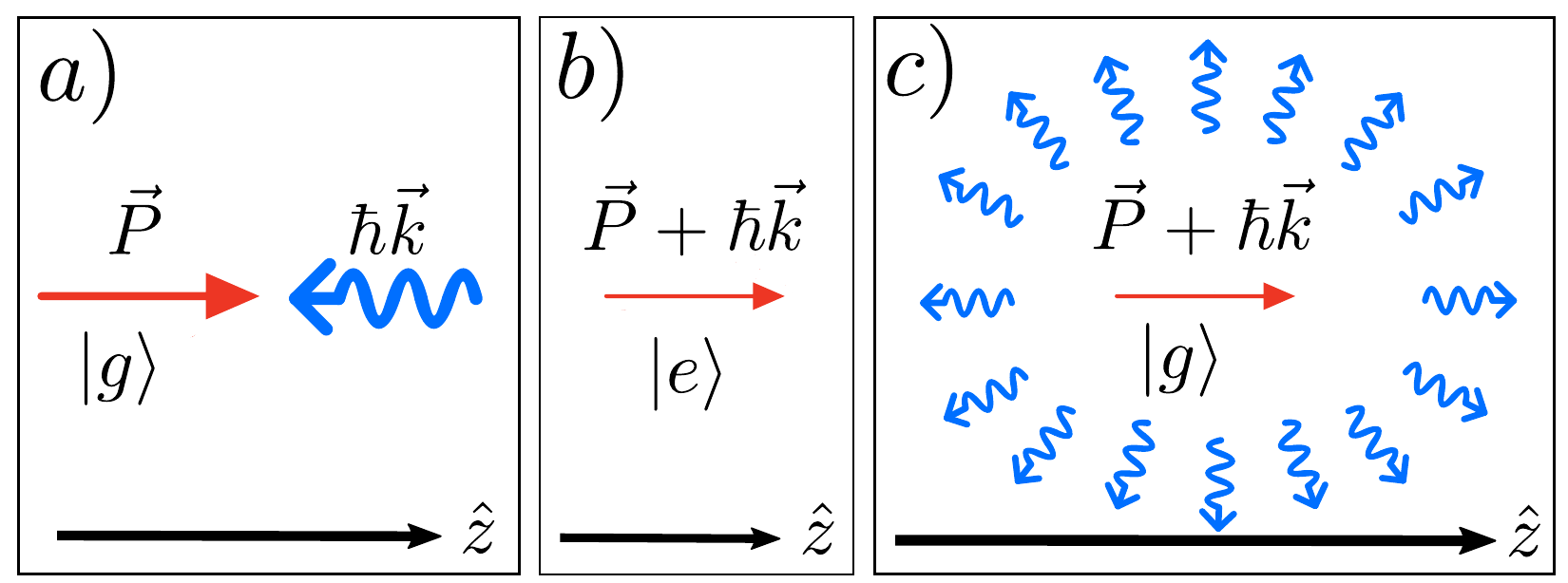}
    \caption{In the Doppler cooling process, (a) and (b): an ion absorbs a red-shifted laser coming towards it and (c) emits it in a random direction. The thickness of the arrows represents the magnitude of the momentum. In Fig.~(b), due to the absorption of a photon moving in the opposite direction, the total momentum of the particle will be smaller, hence the thinner arrow. After the isotropic emission of a photon in Fig.~(c), the momentum of the particle on average does not change.}
    \label{fig:doppler_effect}
\end{figure}
Now, consider the emission process in which the excited ion emits a photon in a random direction (Fig.~\ref{fig:doppler_effect}). Once again, the energy and momentum conservation yields
\begin{equation}
    \omega_0 = \omega_{\rm em} - \vec{v'} \cdot \vec{k}_{\rm em} + \frac{\hbar k^2}{2M} \;,\label{eq:doppler_emission}
\end{equation}
where the subscript `em' represents emission process and $|\vec{k}_{\rm em}| = |\vec{k}|$. We neglect the recoil shift again to get $\omega_{\rm em} = \omega_0 + \vec{v'}\cdot
\vec{k}_{\rm em}$. But the wave vector of the emitted photon $\vec{k}_{\rm em}$ is in a random direction. Hence, averaging over a number of scattering processes we get $\langle \hbar \vec{v'}\cdot \vec{k}_{\rm em}\rangle = 0$. So, on average the energy change of the photon per scattering event is
\begin{equation}
    \hbar \Delta \omega = \langle \hbar(\omega_{\rm em} - \omega_{\rm abs}) \rangle = -\hbar \vec{v}\cdot \vec{k}\;.
\end{equation}
This gain $\vec{v}\cdot \vec{k}$ in photon energy is equal to the loss of the ion’s kinetic energy ($E_K$), i.e., $\Delta E_K = \hbar \vec{v}\cdot \vec{k} < 0$. If we interpret temperature as $\frac{3}{2}k_B T = E_K$, this loss of kinetic energy leads to cooling the ions moving towards the laser.

This process of cooling down ions does not go on indefinitely. One limit to this
process can be readily obtained by considering the recoil shift. When the ions have already significantly cooled down, the velocity $\vec{v}$ is small enough so that the third term in Eq.~(\ref{eq:doppler_absorption}) and (\ref{eq:doppler_emission}) cannot be neglected anymore. With the recoil shift incorporated, the change in the ions kinetic energy becomes $\Delta E_K = \hbar \vec{v}\cdot \vec{k} + \hbar^2 k^2/M$. This means that the ions will cool down as long as $\Delta E_K<0$, i.e. $-\hbar \vec{v}\cdot \vec{k} > \hbar^2 k^2/M$ with $\vec{v}\cdot \vec{k}<0$, imposing a lower limit for the temperature known as the \textbf{recoil limit} \cite{dopplerlimit}.

\subsection{Quantum regime: sideband cooling \label{sec:sideband_cooling}}
After Doppler cooling reaches its limit, the energy of the ion is low enough to consider the vibrational degrees of freedom of the QSHO. Usually, after Doppler cooling, the average QSHO number state of the ion is $\langle n \rangle \sim 10$. However, as stated earlier, our goal is to reach the ground state of the harmonic oscillator to initialize the qubit.  This requires further cooling of the system, which is generally achieved by a process called \textbf{sideband cooling}. For sideband cooling, an internal ionic transition is chosen with linewidth $\Gamma_s$, now small compared to $\omega_z$ (frequency of the QSHO) so that the absorption peak can distinguish frequencies $\omega_0\pm\omega_z$ from $\omega_0$ (as opposed to Doppler cooling where $\Gamma >> \omega_z$). For this reason, a different transition is chosen from that of the Doppler cooling process. We call the lower and higher level of this transition the internal ground state $\ket{g}$ and excited state $\ket{e}$, respectively and denote this transition frequency as $\omega_0$. The free Hamiltonian of the system can be written as
\begin{equation}
    H_0 = \hbar \omega_z a^\dagger a + \frac{\hbar}{2}\omega_0\sigma_z \;,
\end{equation}
where $\sigma_z$ is the Pauli spin operator in Eq.~(\ref{eq:operators}) corresponding to the two-state ion, and $a^\dagger,\,a$ are the raising and lowering operator of the QSHO. Initially the ion is in the state $\ket{g,n}$ with $n\sim 10$. Here, the notation $\ket{g,n}$ means that the ion is in atomic state $\ket{g}$ and in QSHO vibrational state $\ket{n}$. We want to reach the ground state $\ket{g,0}$ to initialize our operational qubit. This is achieved by a repetition of the following transitions:
\begin{equation}
    \ket{g,n}\rightarrow\ket{e,n-1}\rightarrow\ket{g,n-1}\rightarrow\ket{e,n-2}\rightarrow\{\dots{}\}\;,
\end{equation}This set of transitions are implemented by the interaction of ions with a laser, which is considered a classical source of electromagnetic waves. The interaction Hamiltonian is given by
\begin{equation}
    H' = \frac{1}{2}\hbar\Omega\left(\sigma_+ + \sigma_- \right)\left[e^{i(kz-\omega t+\phi)} + e^{-i(kz-\omega t+\phi)}\right]\;, \label{eq:laser-ion-interaction:1}
\end{equation}
where $k,\,\omega$ are the wave vector and frequency of the laser, and $\phi$ is the phase of the laser. $\Omega$ is the coupling strength, $\sigma_+,\,\sigma_-$ are the raising and lowering operator for the internal dof, i.e., $\sigma_+ \ket{g} = \ket{e}$, and $\sigma_-\ket{e} = \ket{g}$. Such Hamiltonian is standard in semiclassical descriptions of light-matter interactions \cite{book:haken}, and models the simplest type of exchange between an atom and a specific component of frequency $\omega$ of the electromagnetic field: the usual sinusoidal oscillating factor is represented by the sum $\frac12\left(e^{i\dots{}}+e^{i\dots{}}\right)\equiv \cos\dots{}$.

Note that we have not quantized the electromagnetic waves. However, the position of the atom ($z$) is quantized since $z = \sqrt{\frac{\hbar}{2M\omega_z}}(a+a^\dagger)$ is determined by the QSHO potential. We define $\eta= k\sqrt{\frac{\hbar}{2M\omega_z}}$, a parameter known as the Lamb-Dicke parameter. We now execute the following steps in consecutive order:
\begin{itemize}
    \item Write $e^{\pm i\eta (a+a^\dagger)} \approx 1\pm i\eta(a+a^\dagger)$, as the Lamb-Dicke parameter is usually small in these experiments \cite{footnote:lamb-dicke}.
    \item Go to the interaction picture by substituting $\sigma_- \rightarrow \sigma_- e^{-i\omega_0 t}$ and its adjoint for $\sigma_+$. Use the rotating wave approximation (RWA) to neglect highly oscillating terms (because on an appreciable timescale, the fast oscillations average out to zero), where the addition of two frequencies $\omega$ and $\omega_0$ appears in the phase.
    \item Finally write interaction picture operators for the vibrational dof $a \rightarrow e^{-i\omega_z t}$.
\end{itemize}

\begin{figure}[ht]
    \centering
    \includegraphics[width=0.8\linewidth]{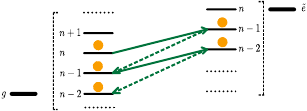}
    \caption{ The figure shows the schematics of sideband cooling. Solid and dashed lines represent stimulated absorption and spontaneous emission processes, respectively.}
    \label{fig:sideband_transition}
\end{figure}

After following these steps, the interaction picture Hamiltonian can be written as
\begin{align}
    H_I = \frac{1}{2}\hbar\Omega\left(\sigma_+ e^{-i(\Delta t - \phi)} + \sigma_- e^{i(\Delta t - \phi)}\right) + &\frac{1}{2}\hbar \eta \Omega \left(a\sigma_+e^{-i(\Delta + \omega_z) t + i\tilde{\phi}} + a^\dagger\sigma_- e^{i(\Delta + \omega_z) t - i\tilde{\phi}} \right) \nonumber \\ + &  \frac{1}{2}\hbar \eta \Omega \left(a^\dagger \sigma_+e^{-i(\Delta - \omega_z) t + i\tilde{\phi}} + a\sigma_- e^{i(\Delta - \omega_z) t - i\tilde{\phi}} \right)\,,
\label{eq:laser-ion-interaction:3}
\end{align}
where $\Delta = \omega -\omega_0$, $\tilde{\phi} = \phi+\pi/2$. The terms in the Hamiltonian are grouped into three contributions. By choosing the laser detuning $\Delta$, one can achieve resonance frequency for each one of the three terms and at resonance the other contributions can be ignored. So, there are three different resonance frequencies corresponding to the three terms in the Hamiltonian. In this section, we focus on only one of the resonances. If we choose the detuning as $\Delta = -\omega_z$, the second term gives the resonant time-independent Hamiltonian

\begin{equation}
    H_I^{(\rm rsb)} \approx \frac{1}{2}\hbar\eta\Omega \left(a \sigma_+ e^{i\tilde{\phi}} + a^\dagger \sigma_- e^{-i\tilde{\phi}} \right)\;.
    \label{eq:rsb_hamiltonian}
\end{equation}
Since the Hamiltonian $H_I^{(\rm rsb)}$ is time-independent, the time-evolution operator becomes $\mathcal{U}_I^{(\rm rsb)}(t) = e^{-iH_I^{(\rm rsb)}t/\hbar}$. As stated earlier, after the Doppler cooling, the ion is in general in a state $\ket{g,n}$ with $n\sim 10$. Applying the time evolution operator to such a state $\ket{g,n}$ yields
\begin{equation}
    \mathcal{U}_I^{(\rm rsb)}(t) \ket{g,n} = \cos(\overline{\omega} t) \ket{g,n} -i e^{i\tilde{\phi}} \sin(\overline{\omega}t) \ket{e,n-1}\label{eq:timeevol}\;,
\end{equation}
where $\overline{\omega} = \eta\Omega\sqrt{n}/2$. The most immediate consequence of (\ref{eq:timeevol}) is that the final state of the ion can be controlled by timing the application of the evolution operator. For example, by choosing to apply the laser pulse for a time $t = \pi/\eta\Omega\sqrt{n}$, the argument of the cosine multiplying the first member becomes $\frac\pi2$, and the whole contribution of the state $\ket{g,n}$ vanishes. The time-evolved state at the end of the interaction process becomes $\ket{e,n-1}$, modulo an overall phase. The ion then spontaneously decays from the $\ket{e,n-1}$ state to the $\ket{g,n-1}$ state. So, at the end of this whole laser-ion interaction the initial state $\ket{g,n}$ reduces to $\ket{g,n-1}$, going one step lower in the QSHO energy level (Fig.~\ref{fig:sideband_transition}). Repeating this process multiple times, one can reach the
ground state of the QSHO with high accuracy, which is now ready for performing computation; thus satisfying criterion 2 of Sec.~\ref{sec:DiVincenzo}. 

The particular choice of frequency $\omega = \omega_0-\omega_z$ for the laser used for the transition $\ket{g,n} \rightarrow \ket{e,n-1}$ is known as \textbf{red sideband resonance} which reflects in the name of the cooling process.

\section{Manipulation of single ions\label{sec:manipulationsingle}}
In the previous section, we saw that by choosing the frequency of the laser, we can use the laser-ion interaction to alter the state of the ion. In particular, we showed that the red sideband resonance is used for a transition of the form $\ket{g,n} \rightarrow \ket{e,n-1}$. In this section, we introduce two more resonant interactions that are used to manipulate an ion during the quantum computing process:
\begin{itemize}
    \item \textbf{Blue sideband resonance}: From Eq.~(\ref{eq:laser-ion-interaction:3}),  if the laser detuning is chosen to be $\Delta = \omega_z$, the third term in the Hamiltonian prevails due to resonance:
    \begin{equation}
        H_I^{(\rm bsb)} \approx \frac{1}{2}\hbar\eta\Omega \left(a^\dagger \sigma_+ e^{i\tilde{\phi}} + a \sigma_- e^{-i\tilde{\phi}} \right)\;.
    \label{eq:bsb_hamiltonian}
    \end{equation}
    The unitary time-evolution operator corresponding to this Hamiltonian induces transitions of the form $\ket{g,n}\leftrightarrow \ket{e,n+1}$, which is known as the blue sideband transition.
    \item \textbf{Carrier resonance}: If the frequency of the laser is chosen to be equal to the frequency of the atomic transition $\omega_0$, i.e. $\Delta = 0$, the resonant Hamiltonian constitutes of the first term in Eq.~(\ref{eq:laser-ion-interaction:3})
    \begin{equation}
         H_I^{(\rm c)} \approx \frac{1}{2}\hbar\Omega \left( \sigma_+ e^{i\phi} + \sigma_- e^{-i\phi} \right)\;.
    \end{equation}
    The time evolution operator in this case does not change the vibrational levels and transitions of the form $\ket{g,n}\leftrightarrow \ket{e,n}$ takes place, which is known as carrier transition.
\end{itemize}

Once we have the fiducial qubit $\ket{g,0}$ at the end of the cooling process, the three resonant processes can be used to create states spanned by $\{\ket{g,0},\ket{g,1},\ket{e,0},\ket{e,1}\}$ because it is possible to go only one step up or down in the QSHO energy levels for $\eta<<1$. Hence, our discussion will now be focused on this subspace of the atom-QSHO Hilbert space. 

The time-evolution operator corresponding to the three transitions can be written as
\begin{equation}
    \mathcal{U}_I = \begin{pmatrix}
    \cos(\beta/2) & -i e^{-i\tilde{\phi}} \sin(\beta/2)\\
    -i e^{i\tilde{\phi}} \sin(\beta/2) & \cos(\beta/2)
    \end{pmatrix} \;,\label{eq:unitary-1qubit-gate}
\end{equation}
where $\beta = \Omega t$ for carrier transitions and $\beta = \eta \Omega t$ for sideband transitions. The phase is $\phi$ instead of $\tilde{\phi}$ for carrier transitions. For each transition, the basis used to write Eq.~(\ref{eq:unitary-1qubit-gate}) consists of the two energy eigenstates that the transition connects:
\begin{itemize}
    \item carrier: $\{\ket{g,0},\,\ket{e,0}\}$,
    \item blue sideband: $\{\ket{g,0},\,\ket{e,1}\}$,
    \item red sideband: $\{\ket{g,1},\,\ket{e,0}\}$.
\end{itemize}

To see how these time-evolution operators are relevant, let us find the action of the carrier resonance operator $\mathcal{U}_I^{\rm (c)}$ on the state $\ket{g,0}$
\begin{equation}
    \mathcal{U}_I^{\rm (c)} \ket{g,0} = \cos(\beta/2) \ket{g,0} - i e^{i\phi} \sin(\beta/2) \ket{e,0}\;. \label{eq:1qubit-carrier-rotation}
\end{equation}
If we now define $\ket{g,0}\equiv \ket{0}$ and $\ket{e,0}\equiv \ket{1}$ as the computational basis, Eq.~(\ref{eq:1qubit-carrier-rotation}) is equivalent to the general 1-qubit state in Eq.~(\ref{eq:qubitstate}) for laser interaction time $t = \theta/\Omega$ and phase $\phi = \varphi - \pi/2$. The general 1-qubit state can be viewed as the outcome of the application of a 1-qubit gate on the fiducial qubit, which implies that any 1-qubit gate can be applied to the fiducial qubit by using the carrier resonance interaction and selecting an appropriate time interval and phase for the laser pulse. From now on, we will drop the subscript $I$ in the time-evolution operator to denote the interaction picture.

\section{Manipulation of multiple ions \label{sec:moreions}}
When more than one ion is present in the trap, an additional phenomenon must be taken into account. Electric charges of the same sign repel one another; therefore, the trapping potential and the repulsive Coulomb potential between ions will compete to determine the ion density. As is often the case when such a competition is present, it is possible to show that an equilibrium state exists. Indeed, for linear traps at usual working temperatures for QC experiments, ions tend to form coherent chain-like structures called \textbf{Coulomb crystals} \cite{cite2}. A full description of the normal modes and the detailed energy levels arising from the study of this structure is outside the scope of this article. A thorough treatment of the case with two ions can be found in \cite{sagawa}. We summarize the results by mentioning that among all its possible collective behaviors, the Coulomb crystal exhibits the lowest energy \textit{center-of-mass} mode, akin to the motion of a single ion, but now involving the whole chain of trapped particles moving in unison.

As we have described in Sec. \ref{sec:manipulationsingle}, the point of our construction so far has been to exploit the interplay between the vibrational and atomic degrees of freedom of a single ion. The ions in this crystal can still be considered as individual atomic qubits, but from a motional point of view, they have to be regarded as a lattice. Consequently, they can no longer change vibrational states individually, and the only allowed energy transitions will necessarily be those associated with the normal modes of the corresponding quantized lattice.

This has the following two consequences.
\begin{itemize}
    \item The state of a chain of $N$ ions must be described as the tensor product of $N$ individual atomic qubits and just \textbf{one global} vibrational qubit:
\begin{equation}
   \left\{\ket{\mathfrak a_1, \nu_1},\dots{},\ket{\mathfrak a_n, \nu_n}\right\}\rightarrow
   \left\{ \ket{\mathfrak a_1\dots \mathfrak a_n, \nu} \right\},\qquad
    \quad(\mathfrak a_j \in\{g,e\},\,\,\nu \in \{0,1\})\;.
\label{eq:ionchainstate}
\end{equation}

    \item Red and blue sideband transitions will now change the vibrational state of \textbf{the whole chain}, although they still operate on \textit{individual} atomic states of the ions. For example, a red sideband transition applied to the second ion of a two-ion system yields
\begin{equation}
\mathcal{U}_2^{\rm (rsb)}(t)| g_1, g_2,\boldsymbol{1} \rangle 
 =\cos(\beta/2)\left| g_1, g_2,\boldsymbol{1} \right\rangle 
 - i e^{i\tilde{\phi}}\sin(\beta/2)\left| g_1, e_2,\boldsymbol{0} \right\rangle \,.\label{eq:unitaryevolution_carrier_moreions}
\end{equation}
It is useful to state explicitly that the energy exchanged in a vibrational transition of the lattice is now $N$ times larger than the one formerly involved in single-ion transitions, as $N$ ions have to change their motion at once.
\end{itemize}

The fact that the entire lattice is now affected by a vibrational transition implies a sort of long-range interaction between the ions in the crystal. This paves the way for creating \emph{entanglement} between ions, which will be the foundation for the implementation of the $CNOT$ gate in Sec.~\ref{sec:QIusingTI}. Current linear traps can host approximately $\sim10^2$ ions, but proposals to scale the system (to satisfy criterion 1b in Sec.~\ref{sec:DiVincenzo}) by building trap networks are currently under development. In such setups, photons are used to transfer information between traps \cite{conclusions:5,conclusions:6,conclusions:7} and more recently, shuttling ions \cite{Pino2021}.

\section{Ion readout\label{sec:ion_readout}}
Once a qubit is prepared, criterion 5 in Sec.~\ref{sec:DiVincenzo} requires the ability to \textbf{reliably} read its status. We know that we can't have an exhaustive description of a quantum system, as a measurement will inevitably collapse it to an eigenstate of the observable. So, we can only adopt a probabilistic approach. A standard way \cite{readout:1} to probe ions involves the selection of an auxiliary excited state $\ket{e'}$, which is distinct from both the ground state $\ket{g}$ and the excited state $\ket{e}$ within the frequency resolution of the instruments. Also, such a state must be \textbf{short-lived}, i.e., decay quickly back to $\ket{g}$.

\begin{figure}[ht]
\centering
\includegraphics[width=0.6\textwidth]{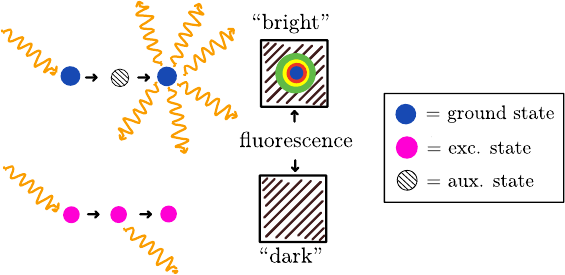}
\caption{Ions interacting with the laser will either fluoresce (if they are in the ground state) or remain inert (if they are in the excited state). The emitted radiation is collected by the photodetector and will result in a bright spot on the detector surface.}
\label{fig:ionreadout}
\end{figure}

The readout procedure then goes as follows. We shine a laser tuned to the transition $\ket{g}\rightarrow \ket{e'}$ on an ion. It will absorb photons (and quickly emit them) if the ion qubit has a component in the ground state. The direction of the emitted photon is random and its radiation can be detected by photodetectors (\textbf{bright} state). On the other hand, an ion in the excited state will be transparent to the radiation (\textbf{dark} state), and no radiation will be detected by the photodetectors. Fig.~\ref{fig:ionreadout} summarizes these concepts.

Shining the laser on an ion is equivalent to measuring its state. So, we can express the process as the collapse of a target state
\begin{equation}
\left| \Psi \right\rangle=\alpha_{\rm bright}\left| g \right\rangle+\alpha_{\rm dark}\left| e \right\rangle
\end{equation}
to either the ground or the excited state, with probabilities $P_{\rm bright}=\left| \alpha_{\rm bright} \right|^2$ and $P_{\rm dark}=\left| \alpha_{\rm dark} \right|^2$. In order to estimate $P_{\rm bright}$ and $P_{\rm dark}$, the preparation and the measurement of the state must be repeated for a sufficiently large number of times. The aforementioned techniques can achieve a precision of 99.99\% and above \cite{readout:2}.

\section{Quantum Computing with Trapped Ions \label{sec:QIusingTI}}
We have shown that a TI system can be accurately prepared in a fiducial state. Such a system has been proven experimentally to be scalable \cite{Pogorelov2021} up to about hundred ions, can be manipulated using laser-ion interaction, and the final state can be effectively read out.  We have also demonstrated that any 1-qubit gate can be applied to the initial state by using a laser pulse with a chosen phase, duration, and carrier transition frequency. However, it remains to be seen if any multiple-qubit gate can be applied to the system. As discussed in Sec.~\ref{sec:universalgates}, any multiple-qubit gate can be obtained using 1-qubit gates and the 2-qubit $CNOT$ gate. In this section, we discuss the process for implementing the $CNOT$ gate. We will show that a $CNOT$ gate can be built using \textbf{Hadamard} and \textbf{Controlled-Z} gates. Before that, we will discuss how to implement these two building blocks below. As before, we define $\ket{g,0}\equiv\ket{0}$, and $\ket{e,0}\equiv \ket{1}$ as the computational basis \cite{operational_basis}. Although both the atomic state and the vibrational states can be used as separate qubits, typically the vibrational qubits are only considered as auxiliary qubits (for intermediate computation) and not as measurable qubits because they are difficult to measure. 

\subsection{Hadamard gate}
The Hadamard gate $H$ is a unitary operation defined by the following action on the computational basis:
\begin{equation}
    H\ket{0}=\frac{\ket{0}+\ket{1}}{\sqrt{2}},~~~~ H\ket{1}=\frac{\ket{0}-\ket{1}}{\sqrt{2}}\;.
\end{equation}
In the $\{\ket{0},\ket{1}\}$ basis the gate is represented by
\begin{equation}
    H = \frac{1}{\sqrt{2}}\begin{pmatrix}
    1 ~& ~1\\
    1 ~& -1
    \end{pmatrix} \;.\label{eq:hadamard-matrix}
\end{equation}
Using Eq.~(\ref{eq:unitary-1qubit-gate}) for a carrier transition, we can check that if we use a laser pulse with $\beta = \frac{\pi}{2}$ and $\phi = -\pi/2$ followed by another laser pulse with $\beta = \pi$ and $\phi=\pi$, we get Eq.~(\ref{eq:hadamard-matrix}) upto an overall phase, i.e., $H = \mathcal{U}^{(c)}(\beta=\pi,\phi=\pi)\; \mathcal{U}^{(c)}(\beta=\pi/2,\phi=-\pi/2) $, modulo a phase factor.

\subsection{Controlled-$Z$ gate}
The controlled-$Z$ ($CZ$) gate is a \textbf{two-qubit gate}, which performs a task on a target qubit depending on the control qubit, similar to a $CNOT$ gate. Its action can be described as:
\begin{itemize}
    \item does nothing to the target qubit if the control qubit is $\ket{0}$,
    \item changes the phase of the target qubit by $\pi$ (or equivalently flips the sign) if the control qubit is $\ket{1}$. 
\end{itemize}
Since $CZ$ is a two-qubit gate, we need two labels for the atomic degrees of freedom of the two ions. The basis for the two-qubit Hilbert space is then $\{\ket{gg},\ket{ge},\ket{eg},\ket{ee}\}$, where the first label denotes the first ion (control qubit) and the second label denotes the second ion (target qubit). In the matrix notation, the $CZ$ gate can be expressed as
\begin{equation}
 CZ \coloneqq  ~~ \begin{blockarray}{cccccc}
 \ket{gg}  & \ket{ge} & \ket{eg}  & \ket{ee}\\
\begin{block}{(cccc)cc}
  1 &  0 & 0 & 0 & ~ & \ket{gg}\\
  0 &  1 &  0 & 0 & ~ & \ket{ge} \\
  0 & 0  &  1 & 0 & ~ & \ket{eg}\\
  0 & 0  &  0 & -1 & ~ & \ket{ee}\\
 \end{block}
\end{blockarray}\;.
\label{eq:CZdefinition}
\end{equation}
To implement the $CZ$ gate we need to construct two more one-ion gates that are described below. \\

\noindent\textbf{Swap gate between atomic and vibrational states} (SWAP$_{av}$):
In this case, we consider both the atomic and the vibrational state of an ion. The basis for this space is $\{\ket{g,0},\ket{g,1},\ket{e,0},\ket{e,1}\}$. The action of the SWAP$_{av}$ gate is represented by the matrix 
\begin{equation}
\text{SWAP$_{av}$} \coloneqq ~~ \begin{blockarray}{cccccc}
\ket{g,0} & \ket{g,1} & \ket{e,0} & \ket{e,1}  \\
\begin{block}{(cccc)cc}
 1 &  0 & 0 & 0 & ~ & \ket{g,0}\\
 0 &  0 &  1 & 0 & ~ & \ket{g,1}\\
 0 & -1  &  0 & 0 & ~ & \ket{e,0}\\
 0 & 0  &  0 & 1 & ~ & \ket{e,1}\\
\end{block}
\end{blockarray}\label{eq:swap_gate}
\end{equation}
with the only non-trivial actions given by $\ket{g,1}\rightarrow - \ket{e,0}$, and $\ket{e,0}\rightarrow \ket{g,1}$. The SWAP$_{av}$ gate can be implemented by a laser detuned to the red sideband. Using Eq.~(\ref{eq:unitary-1qubit-gate}) for the red sideband transition (in the $\{\ket{g,1},\ket{e,0}\}$ basis), the SWAP$_{av}$ gate can be performed by choosing a laser pulse with $\beta = \pi$ and $\tilde{\phi} = 3\pi/2$. The inverse of the gate can be implemented by changing the phase of the laser to $\tilde{\phi} = \pi/2$.\\

\begin{figure}[ht]
\includegraphics[width=0.4\textwidth]{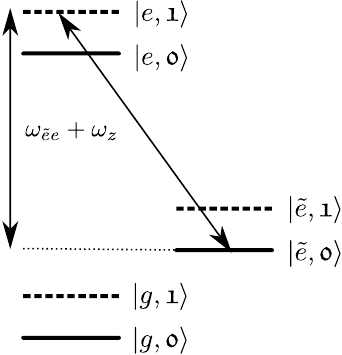}
 \centering
 \caption{Auxiliary level used for the implementation of the $CZ_{av}$ gate. An ion in the atomic ground state remains in the ground state due to the off-resonant frequency of the laser.}
 \label{fig:auxiliary_level}
\end{figure}

\noindent \textbf{$CZ$ gate between atomic and vibrational states} ($CZ_{av}$): This is a controlled-$Z$ gate acting on the atomic and vibrational states of a single ion, where the atomic dof acts as the control qubit and the vibrational dof acts as the target qubit. In the single-ion basis $\{\ket{g,0},\ket{g,1},\ket{e,0},\ket{e,1}\}$, the gate is represented by
\begin{equation}
    CZ_{av}\coloneqq~~ \begin{blockarray}{cccccc}
     \ket{g,0} & \ket{g,1} & \ket{e,0} & \ket{e,1}  \\
    \begin{block}{(cccc)cc}
     1 &  0 & 0 & 0 & ~ & \ket{g,0}\\
     0 &  1 &  0 & 0 & ~ & \ket{g,1}\\
     0 &  0  &  1 & 0 & ~ & \ket{e,0}\\
     0 & 0  &  0 & -1 & ~ & \ket{e,1}\\
    \end{block}
    \end{blockarray} \label{eq:CZ-av-gate}
\end{equation}
with the nontrivial action $\ket{e,1}\rightarrow -\ket{e,1}$. To implement this gate, an auxiliary level $\ket{\tilde{e},0}$ of the ion is found that can be reached only from the $\ket{e,1}$ state by means of a laser tuned to $\omega_{\tilde ee}+\omega_z$ (Fig.~\ref{fig:auxiliary_level}). This is a blue sideband transition in the $\{\ket{\tilde{e},0},\ket{e,1}\}$ basis with the former being the lower energy state replacing $\ket{g,0}$ in the usual description. Using Eq.~(\ref{eq:unitary-1qubit-gate}) for the blue sideband resonance in this new basis, we get $\ket{e,1}\rightarrow -\ket{e,1}$ by choosing $\beta=2\pi$. Since we are using an auxiliary level for the transition, the other original states are left unchanged, giving the only nontrivial change in (\ref{eq:CZ-av-gate}).

We are now ready to implement the $CZ$ gate for two ions. The most generic state of a two-ion system in the vibrational ground state is given by
\begin{equation}
\left| \Psi \right\rangle=\alpha\left| gg,0 \right\rangle+\beta\left| ge,0 \right\rangle+\gamma\left| eg,0 \right\rangle+\delta\left| ee,0 \right\rangle \,,\label{eq:CZinitial}
\end{equation}
where the coefficients $\alpha,\beta,\gamma,\delta$ satisfy the normalization condition.
The $CZ$ gate can be implemented on this state by following the procedure below step-by-step (bold fonts in the following steps highlight the changes in the state $\ket{\Psi}$).
\begin{itemize}
\item Use the SWAP$_{av}$ gate on the second ion:
\begin{equation}
\alpha\left| gg,0 \right\rangle+\beta\left| g\boldsymbol{g},\boldsymbol{1} \right\rangle+\gamma\left| eg,0 \right\rangle+\delta\left| e\boldsymbol{g},\boldsymbol{1} \right\rangle \,.
\end{equation}
\item Apply a $CZ_{av}$ gate to the first ion:
\begin{equation}
\alpha\left| gg,0 \right\rangle+\beta\left| gg,1 \right\rangle+\gamma\left| eg,0 \right\rangle-\delta\left| \boldsymbol eg,\boldsymbol{1} \right\rangle \,. \end{equation}
\item Use the inverse of the SWAP$_{av}$ gate on the second ion:
\begin{equation}
\alpha\left| gg,0 \right\rangle+\beta\left| g\boldsymbol e, \boldsymbol{0} \right\rangle+\gamma\left| eg,0 \right\rangle-\delta\left| e\boldsymbol e,\boldsymbol{0} \right\rangle \,.\label{eq:CZfinal}
\end{equation}
\end{itemize}
Notice that after the set of operations described above, the vibrational qubit returns to the ground state (where it started from). The vibrational qubit can now be dropped from the description of the initial and the final step, which gives the effective transformation
\begin{equation}
\alpha\left| gg \right\rangle+\beta\left| ge\right\rangle+\gamma\left| eg \right\rangle+\delta\left| ee \right\rangle    \longrightarrow \alpha\left| gg \right\rangle+\beta\left| ge \right\rangle+\gamma\left| eg \right\rangle-\delta\left| ee \right\rangle\;. \label{eq:CZ-transformation}
\end{equation}
From Eq.~(\ref{eq:CZ-transformation}), it is now apparent that the steps described above indeed implement the $CZ$ gate on a two-ion system.

\subsection{$CNOT$ gate}
We are finally ready to implement the $CNOT$ gate. One can verify that the $CNOT$ gate can be decomposed as
\begin{equation}
  CNOT\coloneqq ( \mathbb{I}\otimes H)\, CZ \, (\mathbb{I}\otimes H) \label{eq:CNOT_decomposition}
\end{equation}
by simple matrix multiplication. Here, $\mathbb{I}\otimes H$ means that the first ion remains unchanged (hence the identity operation), and the Hadamard gate acts on the second ion. Using the matrix representation, the $CNOT$ gate can then be written as:
\begin{equation}
 CNOT \coloneqq  ~~ \begin{blockarray}{cccccc}
 \ket{gg}  & \ket{ge} & \ket{eg}  & \ket{ee}\\
\begin{block}{(cccc)cc}
  1 &  0 & 0 & 0 & ~ & \ket{gg}\\
  0 &  1 &  0 & 0 & ~ & \ket{ge} \\
  0 & 0  &  0 & 1 & ~ & \ket{eg}\\
  0 & 0  &  1 & 0 & ~ & \ket{ee}\\
 \end{block}
\end{blockarray}\;.
\label{eq:CNOT}
\end{equation}

We have previously described how to build the Hadamard gate and the $CZ$ gate using laser pulses. Using these building blocks, one can now construct the $CNOT$ gate by following the decomposition in Eq.~(\ref{eq:CNOT_decomposition}):
\begin{itemize}
    \item Apply a Hadamard gate on the second ion.
    \item Apply a $CZ$ gate with the first ion as the control qubit and the second ion as the target qubit.
    \item Apply another Hadamard gate on the second ion.
\end{itemize}

It is worth noting that Monroe \textit{et al.} first realized the $CNOT$ gate in a trapped-ion system in 1995 by using the atomic and vibrational states of a Berillyum ion \cite{CNOT}.

\subsection{Creation of Bell states}
The ultimate advantage of quantum computation lies in using the power of entangled qubits (non-classical correlation between qubits). As a final step to demonstrate the viability of trapped ions as a quantum computing platform, we will show that entangled ions can be created and used to execute quantum algorithms in a lab. By applying the operation $ CNOT \, (H\otimes \mathbb{I})$ on the two-ion ground state one can create the maximally entangled Bell state $\ket{\Phi^+}$ 
\begin{align}
     CNOT \, (H\otimes \mathbb{I})\ket{gg} &=  CNOT\frac{\ket{gg}+\ket{eg}}{\sqrt{2}} \nonumber\\
     &= \frac{\ket{gg}+\ket{ee}}{\sqrt{2}} \equiv \ket{\Phi^+}\,.
\end{align}
In fact, all four Bell states can be created using the $CNOT$ gate and one-qubit rotation gates. The Bell states are used as resources in all quantum algorithms that have the potential to provide an advantage over existing classical algorithms.  

\section{Concluding remarks\label{sec:Conclusion}}
Throughout the article, we have shown that trapped ion systems satisfactorily meet the Di Vincenzo criteria for their use as a platform for quantum computing. In short,
\begin{enumerate}
    \item {\it Criteria 1(a) and 2}: Trapped ion qubits are robust and can be initialized to high accuracy.
    \item {\it Criterion 3}: The ions have a high coherence time ($\sim 10^1$ s) to gate time ($\sim 10^{-6}$ s) ratio, giving enough time to perform computations on the system before the effects of decoherence creep in.
    \item {\it Criteria 4 and 5}: It is possible to implement a universal set of gates and read individual ions by carefully tuned laser pulses.
\end{enumerate}
In addition, trapped ions can be entangled easily despite being physically far apart, due to their manipulation via laser pulses. This is a significant advantage over superconducting qubits, making trapped ions a desirable platform for simulating long-range interactions common in quantum many-body physics and high-energy physics.\\
However, scalability (Criterion 1b) to hundreds or thousands of qubits remains a major challenge for trapped ions. Many of the Di Vincenzo criteria break down when the ion chain contains multiple ions. There are two main difficulties:
\begin{enumerate}
    \item {\it Coherence time}: Larger ion chains have shorter coherence times, and the effects of decoherence are much more significant in chains with lengths of 100 ions or more.
    \item {\it Readout}: As the chain gets longer, it becomes harder to read individual ions with high accuracy because the precision required for the laser pulse parameters to measure individual ions exceeds current technology capabilities.  
\end{enumerate}
As of this writing, no major advancement related to scalability has been reported. However, recently some promising techniques have been proposed to overcome the limitations of the current approaches \cite{Pino2021}. On a partially related note, clever manipulation schemes aimed at coupling systems of $N>2$ trapped ions have been presented \cite{katz2023, Andrade2022}.

Apart from scalability challenges, the absolute gate time for trapped ions is longer ($\sim 10^{-6}$ s) than that for superconducting qubits ($10^{-9}$ s). Therefore, performing the same computation on a trapped ion platform would take a much longer time and present a challenge to achieve quantum advantage.

Although these problems may seem insurmountable, physicists are continuously developing more creative ways to address them. For a recent and comprehensive review, we encourage readers to see Ref.~\cite{Bruzewicz2019}. While scalability and decoherence issues are yet to be resolved, this is what makes this field exciting and challenging. Until then, we remain in the era of noisy intermediate-scale quantum (NISQ) devices, and it is difficult to predict which platform will emerge victorious, if any. Nevertheless, physicists have already used NISQ devices to simulate physical problems and provided proof-of-concept implementations (for example, see Refs.~\cite{preskillNISQ,Zoller-Troyer,quantum-chemistry}), and we look forward to further developments in the near future.

\section{Acknowledgments} 
We'd like to thank Professor Jose Goity for kindly giving us access to his quantum computing lecture notes, and valuable feedback on an early version of this review. One of us (C.R.O.) would like to thank K. Hazzard, G. Pagano, J. Kuno, and M. O. Scully for many illuminating conversations on atomic, molecular, and optical physics (AMO), and to the latter, for fruitful collaborations in the interception of quantum gravity and quantum optics. This material is based upon work supported by the Air Force Office of Scientific Research (AFOSR) under Grant No. FA9550-21-1-0017 (C.R.O., A.C., F.B.). C.R.O. was partially supported by the Army Research Office (ARO), grant W911NF-23-1-0202.

{}

\begin{thebibliography}{}
\bibitem{dopplerlimit}
The recoil limit is usually less than the Doppler limit which arises due to the non-zero linewidth of the absorption maxima. The Doppler limit takes place when the Doppler shift $\vec{v}\cdot \vec{k}$ becomes comparable to $\Gamma$ and in most cases provides the lowest temperature achievable by Doppler cooling. For more details see Ref. \cite{dopplercooling:1}.
\bibitem{footnote:lamb-dicke}
This small Lamb-Dicke parameter regime has another interpretation. We notice that $\eta^2 = \frac{\hbar^2k^2/2M}{\hbar \omega_z}$, which is the ratio of the recoil energy and the QSHO transition energy. For $\eta<<1$, the recoil energy is small compared to the transition energy which ensures that spontaneous emission occurs only between levels $\ket{e,n}\rightarrow \ket{g,n}$, i.e. at the carrier frequency.
\end{thebibliography}
\end{document}